\documentclass[10pt,twocolumn,aps,english,pra,superscriptaddress,showpacs]{revtex4}
\usepackage{amsfonts}
\usepackage[T1]{fontenc}
\usepackage[latin9]{inputenc}
\usepackage{amsmath}
\usepackage{amssymb}
\usepackage{graphicx}
\usepackage{babel}
\usepackage{booktabs}
\usepackage{tabu}
\usepackage{subfigure}
\usepackage{physics}
\usepackage{xr}
\makeatletter
\@ifundefined{textcolor}{}
{
 \definecolor{BLACK}{gray}{0}
 \definecolor{WHITE}{gray}{1}
 \definecolor{RED}{rgb}{1,0,0}
 \definecolor{GREEN}{rgb}{0,1,0}
 \definecolor{BLUE}{rgb}{0,0,1}
 \definecolor{CYAN}{cmyk}{1,0,0,0}
 \definecolor{MAGENTA}{cmyk}{0,1,0,0}
 \definecolor{YELLOW}{cmyk}{0,0,1,0}
 }
\makeatother

\begin{document}

\title{Controlled quantum search on structured databases}

\author{Yunkai Wang}
\affiliation{Kuang Yaming Honors School, Nanjing University, Nanjing, Jiangsu 210023,
China}
\affiliation{Department of Physics, University of Illinois at Urbana-Champaign, 1110 West Green Street, Urbana, Illinois 61801, USA}

\author{Shengjun Wu}
\affiliation{Kuang Yaming Honors School, Nanjing University, Nanjing, Jiangsu 210023,
China}
\affiliation{Collaborative Innovation Center of Advanced Microstructures, National Laboratory of Solid State Microstructure and Department of Physics, Nanjing University, Nanjing 210093, China}

\author{Wei Wang}
\affiliation{Kuang Yaming Honors School, Nanjing University, Nanjing, Jiangsu 210023,
China}
\affiliation{Collaborative Innovation Center of Advanced Microstructures, National Laboratory of Solid State Microstructure and Department of Physics, Nanjing University, Nanjing 210093, China}

\date{\today }

\begin{abstract}

We present quantum algorithms to search for marked vertices in structured databases with low connectivity. Adopting a multi-stage search process, we achieve a success probability close to $100\%$ on Cayley trees with large branching factors. We find that the number of stages required is given by the height of the Cayley tree. At each stage, the jumping rate should be chosen as different values. The dominant term of the runtime in the search process is proportional to $N^{(2r-1)/2r}$ for the Cayley tree of height $r$ with $N$ vertices. We further find that one can control the number of stages by adjusting the weight of the edges in the graphs. The multi-stage search process can be merged into a single stage, and then an optimal runtime proportional to $\sqrt{N}$ is achieved, yielding a substantial speedup. The search process is quite robust under various kinds of small perturbations.

\end{abstract}

\maketitle

\emph{Introduction.}--As one of the important applications of quantum computation, Grover's algorithm is designed for unstructured databases \cite{PhysRevLett.79.325}. In unstructured databases, one can easily move from one vertex to any other vertices for finding the marked (or target) items. However, physical databases usually have structures that might prevent one from moving arbitrarily. Thus, the related spatial search algorithms, aimed to find the marked item, are not so easy to be realized.

Similar to classical random walk, the quantum walk has been used as an algorithmic tool for the spatial search \cite{aharonov1993quantum,venegas2012quantum,kempe2003quantum}. The application of quantum walk has been widely discussed \cite{childs2003exponential,childs2009universal, lovett2010universal,shenvi2003quantum,di2011mimicking,jeong2013experimental,childs2013universal,qiang2016efficient,robens2015ideal,faccin2013degree,solntsev2014generation,flurin2017observing}, and its implementation has also been achieved in different platforms \cite{schmitz2009quantum,zahringer2010realization,xue2009quantum,karski2009quantum,genske2013electric,du2003experimental,perets2008realization,bromberg2009quantum,crespi2013anderson,broome2010discrete,peruzzo2010quantum}.
In particular, as shown by Childs and Goldstone algorithms based on continuous-time quantum walk (CTQW) can be used to solve the spatial search problem on the complete graph, the hypercube and $d$-dimensional periodic lattices \cite{PhysRevA.70.022314}.
Such spatial search algorithms via CTQW have also been applied on several different graphs \cite{PhysRevLett.112.210502,PhysRevLett.114.110503,PhysRevA.93.032305,PhysRevLett.116.100501,novo2015systematic,agliari2010quantum}.
An optimal search time $\sim O(\sqrt{N})$  has been obtained for some of these structured graphs under certain circumstances, and some other properties have also been explored for other graphs \cite{PhysRevLett.114.110503,PhysRevLett.112.210502,agliari2010quantum,PhysRevA.94.022304,PhysRevA.92.032320}. In addition, quantum walk on weighted graphs has also been studied for universal mixing \cite{Quantum.Info.Comput.7.738}, quantum state transfer \cite{PhysRevLett.92.187902}, quantum transport \cite{Sci.Rep.3.2361}, and quantum search \cite{PhysRevA.94.022304,PhysRevA.92.032320}.

Trees, one kind of the most important data structures in computer science, are very convenient for data manipulation due to their low connectivity. As a typical and important case, there have been a number of discussions about quantum algorithm related to trees \cite{childs2003exponential,berry2010quantum,agliari2010quantum,PhysRevA.93.032305, mulken2006coherent, agliari2008dynamics, farhi1998quantum, chisaki2009limit, tregenna2003controlling,carneiro2005entanglement,farhi2008quantum}. However, for trees the marked items are difficult to be found since they can be deeply hidden as the leaf vertices, and none of the reported schemes can achieve an optimal runtime. Moreover, the reported success probability of search on marked vertices is rather low, and practically, decreases when the height of the tree increases.
Is it possible to achieve a spatial search scheme with optimal search time and high success probability for trees, such as the Cayley trees, a typical balanced tree?

In this letter, we try to provide a solution. We present a multi-stage quantum search scheme on the Cayley trees, and find a success probability close to $100\%$ which does not decrease when the height or branching factor of the tree increases. Surprisingly, we also find that proper variations of the edge weights can cause a merging of the stages in the search process, thus resulting in a substantial speedup for the search. With a suitable choice of the edge weights, we can achieve an optimal search time $O(\sqrt{N})$ for the Cayley trees while keeping the success probability close to $100\%$.
Our study indicates that for physically structured databases with low connectivity, quantum search algorithm based on CTQW can still provide a quadratic speedup. Such a behavior of the quantum search independently on the database structure may illustrate certain intrinsic properties of quantum mechanics.

\emph{Multi-stage quantum search on Cayley trees}--We consider a particle that performs a quantum walk on a graph with $N$ vertices, with each vertex corresponding to a basis state in an $N$-dimensional Hilbert space. The quantum walk is governed by the Hamiltonian
\begin{equation}
H=-\gamma L-\ket{a}\bra{a},
\end{equation}
where $\gamma$ is the jumping rate, $L=A-D$ is the graph Laplacian, $A$ is the adjacency matrix of graph (i.e., $A_{ij}$ is the weight of edge between vertices $i$ and $j$), $D$ is the diagonal matrix with $D_{jj}=deg(j)$ (i.e., the total weight of edges connected to vertex $j$), and $\ket{a}$ is the state corresponding to the marked vertex $a$. For convenience, our Hamiltonian $H$ and the time are chosen to be dimensionless.
Since we have no prior information of the location of the marked vertex, the initial state of the particle is chosen as
\begin{equation}
\ket{s}=\frac{1}{\sqrt{N}}\sum_{i=1}^{N}\ket{i}  .   \label{initialsymmetricstate}
\end{equation}
In order to study the evolution of the state, an invariant subspace of the Hamiltonian should be found by grouping the identically evolving vertices together.

\begin{figure}[bt]
\includegraphics[width=0.44\textwidth]{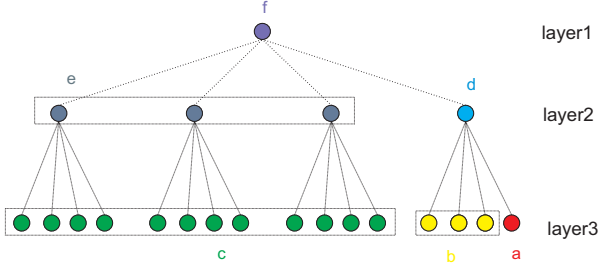}
\caption{A Cayley tree with height $r=2$ and branching factor $M=4$.}
\label{01_1st_tree_group}
\end{figure}

We focus on a Cayley tree with height $r$ and branching factor $M$, which contains one vertex in the first (top) layer, $M$ vertices in the second layer, $\cdots$, and $M^r$ vertices in the $(r+1)$-th (bottom) layer, thus the total number of $(M^{r+1}-1)/(M-1)$ vertices
(see Fig.\ref{01_1st_tree_group}). In our work, we assume that the marked vertex is a leaf vertex.

We first discuss the case for a Cayley tree with height $r=2$ and an arbitrary branching factor $M$. Figure \ref{01_1st_tree_group} shows the case of $M=4$. To study the evolution of the system, we first group the vertices with identical evolution by the same color, and then work out the invariant subspace.
The invariant subspace is spanned by the following $6$ basis states:
$\ket{a}=\ket{a}$ (the marked vertex state), $\ket{b}=\frac{1}{\sqrt{M-1}}\sum_{i\in b} \ket{i}$, $\ket{c}=\frac{1}{\sqrt{M(M-1)}}\sum_{i\in c} \ket{i}$, $\ket{d}$, $\ket{e}=\frac{1}{\sqrt{M-1}}\sum_{i\in e} \ket{i}$, and $\ket{f}$.
A state initially in this subspace evolves only in the subspace, as the off-diagonal terms of the Hamiltonian between the invariant subspace and its orthogonal complement vanish.
Therefore, the effective Hamiltonian in this subspace is
\begin{displaymath}
H
=-\gamma\begin{bmatrix}
-1+\frac{1}{\gamma}& 0 & 0 & 1 & 0 & 0\\
0 & -1 & 0 & \sqrt{M_1} & 0 & 0\\
0 & 0 & -1 & 0 & \sqrt{M} & 0\\
1 & \sqrt{M_1} & 0 & -M_{-1} & 0 & 1\\
0 & 0 & \sqrt{M} & 0 & -M_{-1} & \sqrt{M_1}\\
0 & 0 & 0 & 1 & \sqrt{M_1} & -M\\
\end{bmatrix}
\end{displaymath}
where $M_k=M-k$ with $k=1$ or $-1$.

\begin{figure}[bt]
  \includegraphics[width=0.5\textwidth]{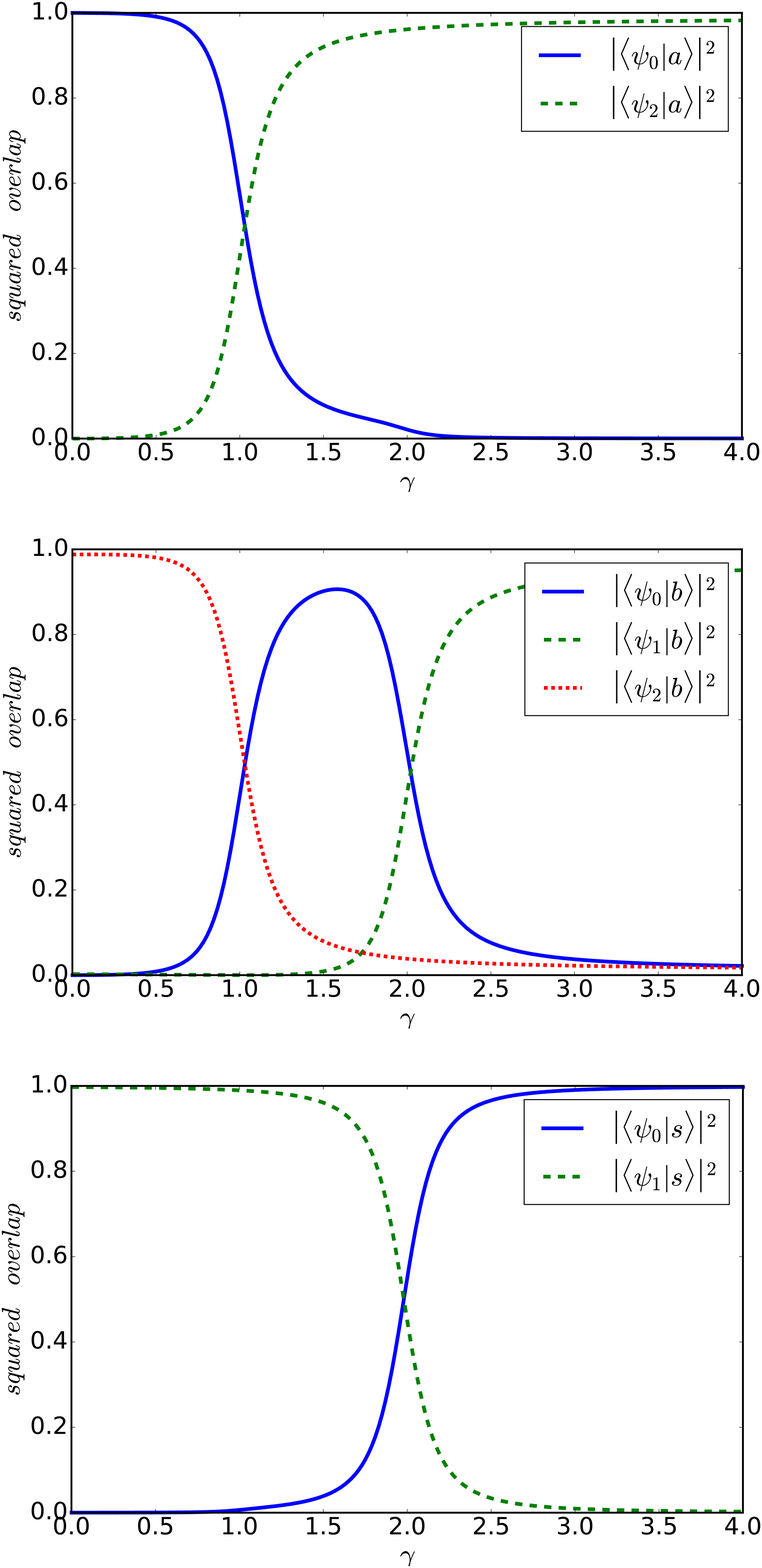}
  \caption{The squared overlaps of basis states with the eigenstates of $H$ for a Cayley tree of height $2$ when $M=100$.}\label{02_overlap_tree}
\end{figure}

The evolution of the particle's state is determined by a Schr\"odinger's equation with the solution $\ket{\psi(t)}=e^{-iHt}\ket{s}=e^{-iHt}\sum_{i=0}^{5}\ket{\psi_i}\bra{\psi_i}\ket{s}$, where $\ket{\psi_i}$ is the $i$-th eigenstate of $H$. Figure \ref{02_overlap_tree} gives the squared overlaps of basis states with the eigenstates of $H$. From Fig.\ref{02_overlap_tree}, we know that when $\gamma=2$, the first two eigenstates of $H$ can be written as $\ket{\psi_{0,1}}\approx (\ket{b}\pm \ket{s})/\sqrt{2}$, and
\begin{equation}
\ket{\psi(t)}= e^{-iHt}\ket{s} \approx \frac{1}{\sqrt{2}}(e^{-iE_{0}t}\ket{\psi_0} + e^{-iE_{1}t}\ket{\psi_1})
\end{equation}
therefore $|\bra{b}\ket{\psi(t)}|^2=(1-cos\Delta E_{10} t)/2$, where $\Delta E_{10}=E_1-E_0$. Hence, the probability amplitude will flow from $\ket{s}$ to $\ket{b}$ within a time $\pi/\Delta E_{10}$. Similarly, when $\gamma=1$, the state of the system oscillates between $\ket{b}$ and $\ket{a}$. In order to make the probability amplitude be accumulated in the marked vertex $a$, a two-stage process is required and works as follows. In the first stage, the system is prepared in the initial state $\ket{s}$, and evolves according to the Hamiltonian with $\gamma=2$ during a time $\pi/\Delta E_{10}$. Then, in the second stage, the system evolves according to the Hamiltonian adjusted by changing the jumping rate $\gamma=2$ to $\gamma=1$ for a time $\pi/\Delta E_{20}$. Finally, a simple projective measurement onto the basis states will reveal the state $\ket{a}$ for the marked vertex $a$ with a very high probability. Here, the special values of $\gamma=2$, $1$ are termed as critical jumping rates $\gamma_c$ approximately. It is worthy to note that when the value of $\gamma$ deviates considerably from that of $\gamma _{c}$, the basis states involved in Figure \ref{02_overlap_tree} become the eigenstates of $H$. Thus, the probability cannot be accumulated in the marked vertex $a$ and the search scheme fails.

For the first stage with $\gamma =2$, we find numerically that $E_{1}-E_{0}=4M^{-3/2}$. Then, the time consumed in this stage is $t=\pi M^{3/2}/4$, and the probability amplitude flows from $\ket{s}$ to $\ket{b}$. For the second stage with $\gamma =1$, we find $E_{2}-E_{0}=2M^{-1/2}$. Then, the time consumed in this stage is $t=\pi M^{1/2}/2$, and the probability amplitude flows from $\ket{b}$ to $\ket{a}$. When $M$ is large enough, e.g., $M=100$, the success probability $|\bra{a}\ket{\psi(t)}|^2$ for finding the marked vertex in the final projective measurement is larger than $99\% $. The effect of $M$ on the success probability is shown in Appendix. \ref{success equal weight}.  The dominant term of the time consumed in the search process is
$t\propto M^{3/2}\propto N^{3/4}$.

\emph{The general cases for the various Cayley trees}--Now, we can explicitly
write down the Hamiltonian in the invariant subspace for Cayley trees with any height $r$, and work out the related results for the cases with the height up to $r=6$. That is, for a Cayley tree with height $r$, we find
that the search process has $r$ stages with different jumping rates $\gamma =r$, $r-1$, $\cdots$, $2$, $1$, and the time consumed in each stage is $t\propto M^{1/2}$, $M^{3/2}$, $\cdots$, $M^{(2r-3)/ 2}$, $M^{(2r-1)/2}$ (i.e., $t\propto N^{1/2r }$, $N^{3/2r}$, $\cdots$, $N^{(2r-3)/2r}$, $N^{(2r-1)/2r}$), respectively. The dominant term of the time consumed in the search process is $t\propto M^{(2r-1)/2}\propto N^{(2r-1)/2r}$, which increases as the height increases.

An interesting observation is that the ground state of the Hamiltonian $\ket{\psi_0}$ changes dramatically when the value of the jumping rate $\gamma$ in the Hamiltonian changes around the related critical jumping rates. The number of these dramatic changes corresponds to the number of searching stages. Furthermore, the probability amplitude flows from the out-most structure to the inner-most structure step by step, indicating that the hierarchical structure of the graphs may be the reason of a multi-stage quantum search. See Appendix. \ref{probability distribution} for more demonstrations. We expect that these properties hold for other hierarchical graphs as well.

In a previous work for the Cayley tree with small branching factor, based on a single-stage search process, the success probability was found to be much less than $100\%$, and decreases as the height increases \cite{agliari2010quantum}. Such a result can be clearly understood from our Fig.\ref{02_overlap_tree} since $\ket{a}$ is dominated by states $\ket{\psi_0}$ and $\ket{\psi_2}$, while $\ket{s}$ is dominated by $\ket{\psi_0}$ and $\ket{\psi_1}$. If the probability amplitude is kept to flow directly from $\ket{s}$ to $\ket{a}$, one should make use of a small component of $\ket{\psi_1}$ in $\ket{a}$, which will lead to a small success probability. Obviously, this component becomes smaller when the number of vertices of the graph becomes larger (see Appendix. \ref{small component}).
Nevertheless, our scheme with a multi-stage search process can yield a success probability close to $100\%$ even for a large branching factor $M$. Moreover, the success probability in our scheme definitely does not decrease when the height $r$ (or the branching factor $M$) of the Cayley tree increases.

\emph{Merging of stages in quantum search}--Our multi-stage quantum search on Cayley trees achieves a high success probability with a runtime $t\propto M^{(2r-1)/2}\propto N^{(2r-1)/2r}$. Can we further improve the search scheme to achieve an optimal runtime $\propto N^{1/2}$ while keeping the high success probability? A positive and even surprising answer is given below. By adjusting the edge weights we can merge our multi-stage search process into a single stage and achieve an optimal search speed with runtime $\propto N^{1/2}$ via this merged single-stage process!

Let us show an example firstly for a Cayley tree with height $2$. The weight of edges between the first and second layer is set as $\omega$, and the invariant subspace is the same as before. Then the effective Hamiltonian in the invariant subspace is written as
\begin{displaymath}
H
=-\gamma\begin{bmatrix}
-1+\frac{1}{\gamma}& 0 & 0 & 1 & 0 & 0\\
0 & -1 & 0 & \sqrt{M_1} & 0 & 0\\
0 & 0 & -1 & 0 & \sqrt{M} & 0\\
1 & \sqrt{M_1} & 0 & -M-\omega & 0 & \omega\\
0 & 0 & \sqrt{M} & 0 & -M-\omega & \omega\sqrt{M_1}\\
0 & 0 & 0 & \omega & \omega\sqrt{M_1} & -\omega M\\
\end{bmatrix}.
\end{displaymath}

\begin{figure}[bt]
  \includegraphics[width=0.45\textwidth]{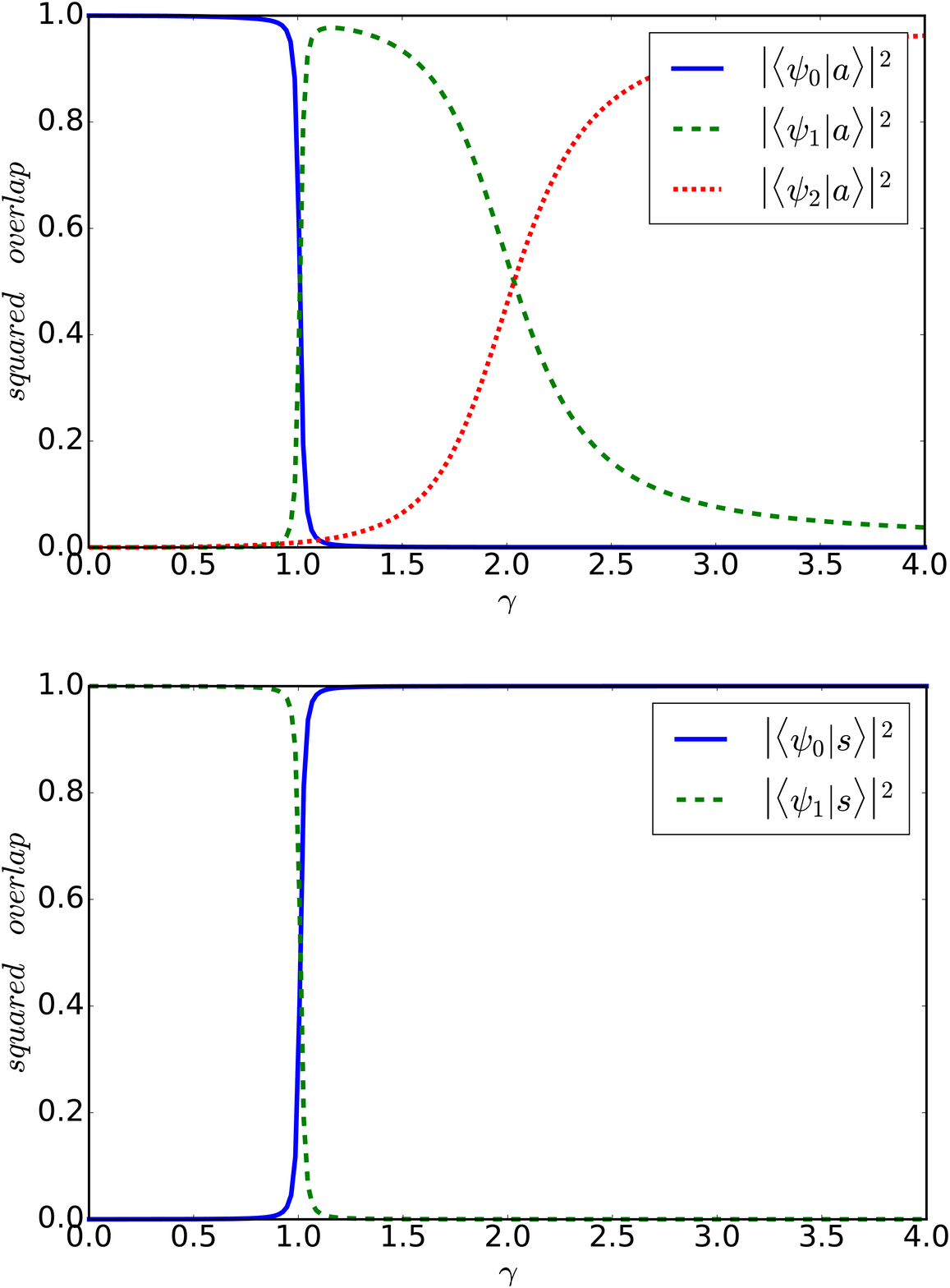}
  \caption{The squared overlaps of basis states with the eigenstates of $H$ for a weighted Cayley tree of height $2$ when $\omega=M$.}\label{03_merging_tree}
\end{figure}

From our numerical calculations, we observe that when the value of $\omega$ increases from $O(M^{1/4})$ to roughly $O(M^{3/4})$, the two stages in the search process are merged gradually, and then become a single stage for the value of $\omega > O(M^{3/4})$. In the following, we present a detailed discussion on the case of $\omega=M$. Figure \ref{03_merging_tree} shows the squared overlaps of basis states with the eigenstates of $H$. When $\gamma=1+1/M$, we find that $E_{1}-E_{0}=2.235M^{-1}$, and the time consumed in the search process is $t=\pi M/2.235\propto N^{1/2}$. Compared to the original two-stage search process, which gives a consumed time $t\propto M^{3/2} \propto N^{3/4}$, a substantial speedup is achieved on the weighted graph.
We also find that the success probability is larger than $97\%$ when $M$ is large enough. It could be more close to $100\%$ with a more delicate choice of the jumping rate $\gamma$.
The influence of $M$ on the success probability is shown in Appendix. \ref{success weighted}. Similar results for a Cayley tree with height $3$ have been obtained (see Appendix. \ref{merging height 3}). The runtime $t \propto N^{5/6}$ in a multi-stage search is speeded up to an optimal runtime $t\propto N^{1/2}$ in the single stage search with different controlled weights.

\begin{figure}[bt]
\includegraphics[width=0.44\textwidth]{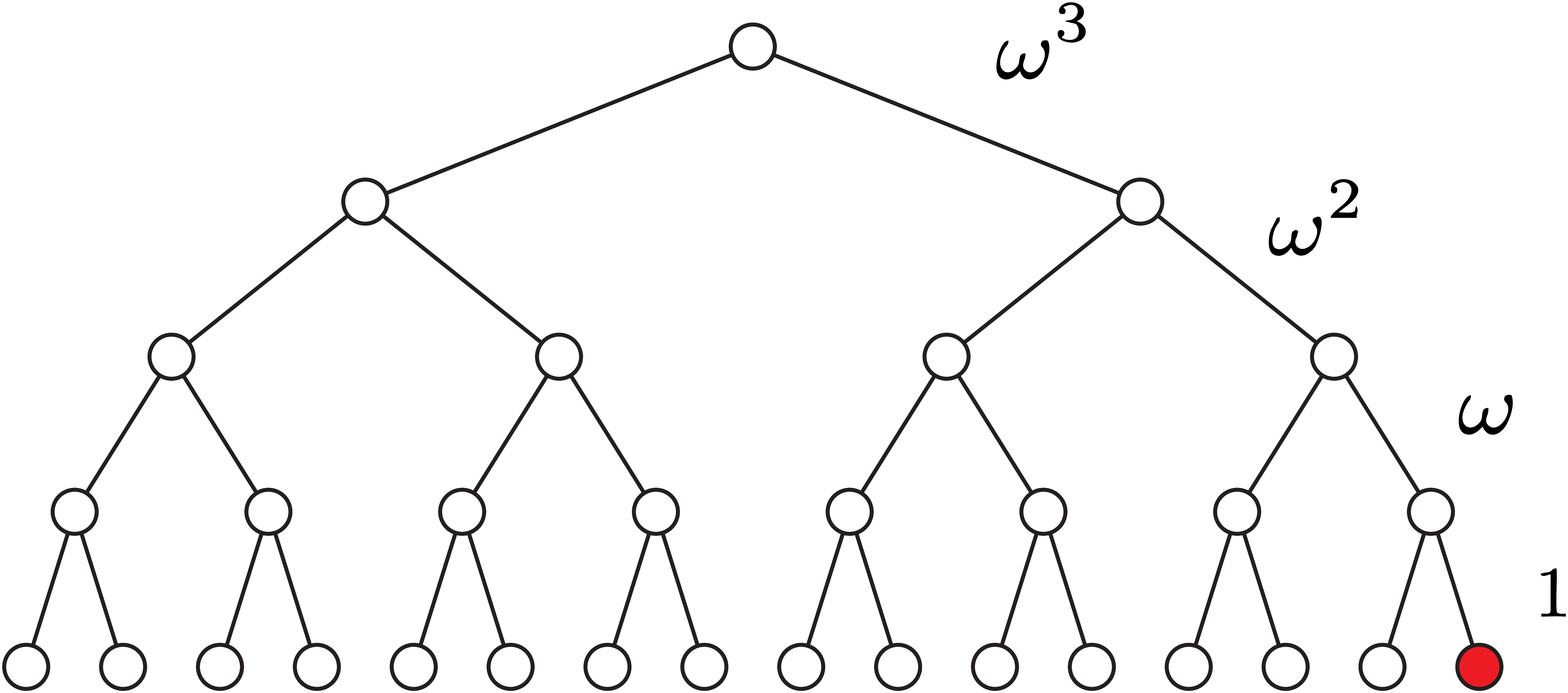}
\caption{Weighted Cayley tree of height $4$ and branching factor $2$. The red vertex is the marked leaf vertex.}
\label{04_balanced_tree}
\end{figure}

Now, let us turn to discuss the case when the value of $M$ is small while $r$ is not small. Interestingly, by properly adjusting the weights of edges between different layers, we can still achieve the optimal search time for small branching factor (e.g. $M=2$) and large height $r$. We set the edge weights to $1$, $\omega$, $\omega^2$, $\omega^3$, $\cdots$, respectively, from bottom up as shown in Figure. \ref{04_balanced_tree}.

As an example, we consider the case when $r=15$, $M=2$ and $\omega=3$. The squared overlaps of basis states with the eigenstates of $H$ are given in Appendix. \ref{overlap small M}. When $\gamma=1.5$ and height varies, we have $E_{1}-E_{0}=1.764 N^{-1/2}$, and find that the time consumed in the search process is $t=\pi N^{1/2}/1.764$. Thus, the optimal search time is still achieved! The success probability is found to be larger than $75\%$. The influence of total number of vertices $N$ on the success probability is shown in Appendix. \ref{success small M}.
This example shows that our scheme with controlled edge weights can still provide an optimal search speed even when the branching factor is small. Compared to a recent scheme on this graph \cite{PhysRevA.93.032305}, which gives a runtime proportional to $N$, our scheme provides a substantial speedup.

Our results show that optimal quantum search can be achieved on intrinsically poorly connected graphs. A Cayley tree is intrinsically poorly connected since it becomes disconnected whichever edge is removed (the joined complete graphs \cite{PhysRevLett.114.110503} is not poorly connected in this sense). It has a low connectivity, irrespective of which measure one uses. According to the definition in \cite{beineke2002average}, the average connectivity of the weighted Cayley tree of height 2 when $\omega=M$ is roughly $4/N$. A discussion on different measures of connectivity can be found in Appendix. \ref{connectivity}.

We notice that a sufficient condition for optimal quantum search using CTQW was proposed in Ref.\cite{PhysRevLett.116.100501}. This condition is not satisfied by the Cayley trees with adjusted edge weights discussed here. However, an optimal quantum search with runtime $O(\sqrt{N})$ is achieved via our scheme. Therefore, it is not a necessary condition for the optimal quantum search. Cayley trees with weighted edges have far richer structure of the Hilbert space in which one can design useful quantum algorithms.

\emph{Conclusion and discussion}--We have discussed quantum search of a marked vertex on Cayley trees with any height and any branching factor. Even though a Cayley tree has a low connectivity, our multi-stage quantum search scheme can achieve a success probability approaching $100\%$ with a large branching factor $M$, and the success probability does not decrease even when the height or branching factor of the tree increases. The height of the Cayley tree directly gives the number of stages needed in the search process. Dominant term of the runtime in the search process is $t\propto M^{(2r-1)/2}\propto N^{(2r-1)/2r}$ for a Cayley tree of height $r$.

We have further found that different stages can be merged into a single stage by controlling the weight of edges between different levels of vertices in the graph. Thus, we can achieve a runtime $t\propto \sqrt{N}$ with a substantial speedup. In other words, by using our search scheme with controllable weights, quantum search of a marked vertex on a Cayley tree with $N$ vertices and a large branching factor can be achieved with a success probability close to $100\%$ in an optimal runtime $t\propto \sqrt{N}$. Our scheme also works for Cayley trees when its branching factor is small while the height is large. In Appendix. \ref{robustness}, we have shown these schemes are also quite robust under small deviations of the jumping rates or small perturbation of the graph structure. Finally, we expect similar results hold for other graphs with hierarchical structures.

\acknowledgments

We thank Yang Gao, Andrew M. Childs and Yunchu Wang for valuable discussions.
This work is supported by the
National Natural Science Foundation of China (Grants No. 11475084 and No.
11275181), and the National Key R\&D Program of China (Grants No. 2016YFA0301801 and No. SQ2017YFJC010006).

\appendix

\section{Effect of branching factor on probability of success}\label{success equal weight}

\begin{figure}[bt]
   \includegraphics[width=0.5\textwidth]{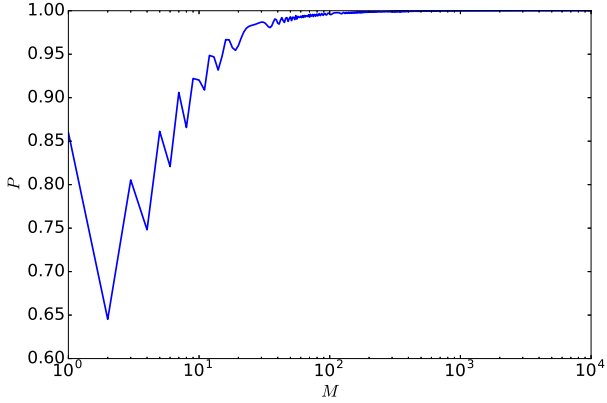}
  \caption{Success probability $P$ versus branching factor $M$ on a Cayley tree of height $2$.}\label{S01_M_success_tree_height2}
\end{figure}

For our two-stage quantum search scheme on a Cayley tree of height 2, we calculate the probability of success for different values of branching factor $M$, and show the effect of $M$ on the success probability in Fig.\ref{S01_M_success_tree_height2}. From the figure, we see that the probability of success for our two-stage quantum search scheme is not affected by the branching factor $M$ as long as $M$ is large enough.

\section{Evolution of probability distribution}\label{probability distribution}

\begin{figure}[bt]

  \includegraphics[width=0.40\textwidth]{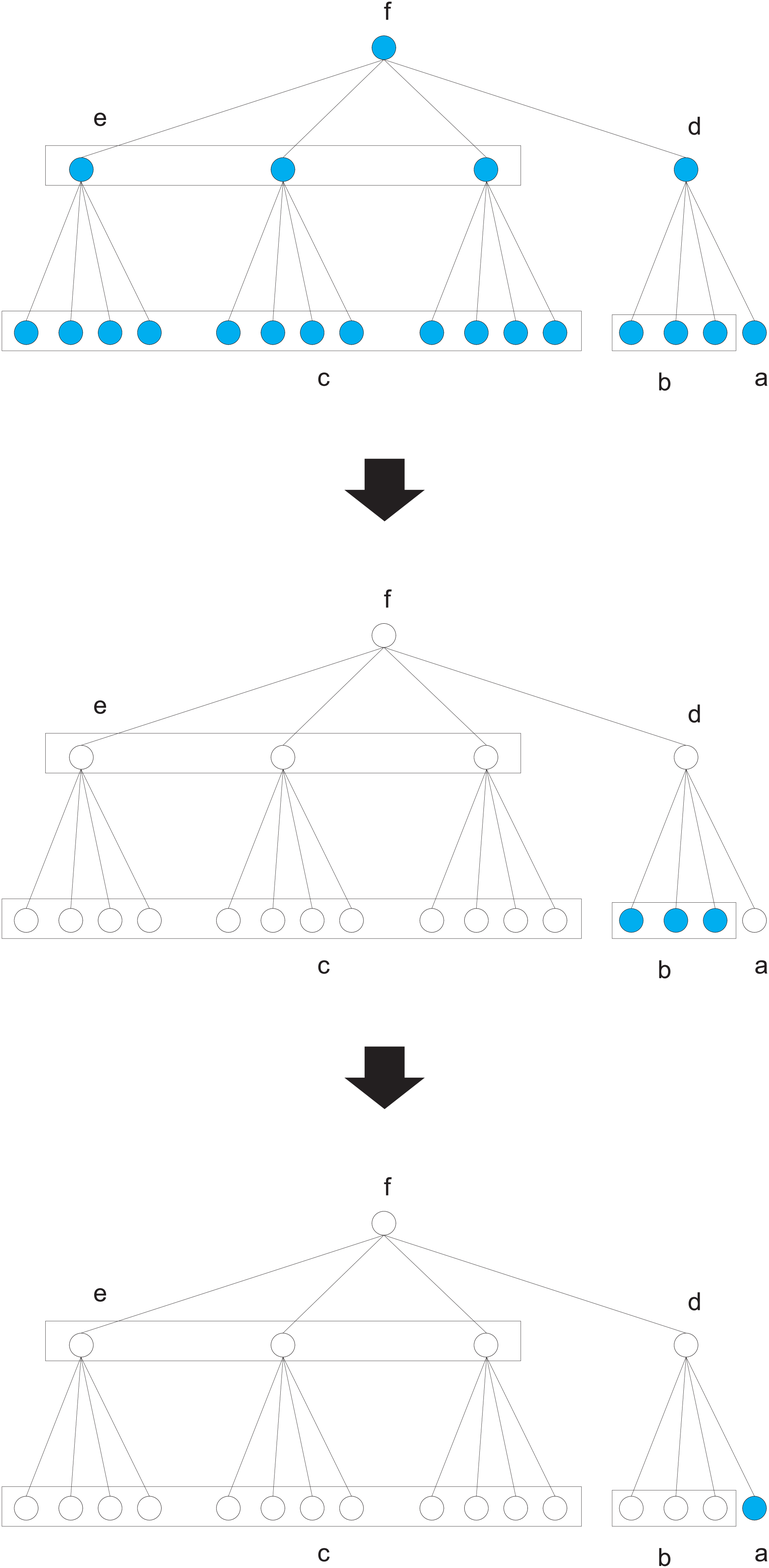}

  \caption{The evolution of the probability distribution on a Cayley tree of height 2.}\label{S02_evolution_tree}
\end{figure}

For our two-stage quantum search scheme on a Cayley tree of height 2, the evolution of probability distribution in the search process is shown in Fig. \ref{S02_evolution_tree}. The dominant part of probability would flow from the out-most structure to the inner-most structure step by step in our scheme.

\section{Small component of $\ket{\psi_1}$ in $\ket{a}$}\label{small component}

\begin{figure}[bt]
  \includegraphics[width=0.5\textwidth]{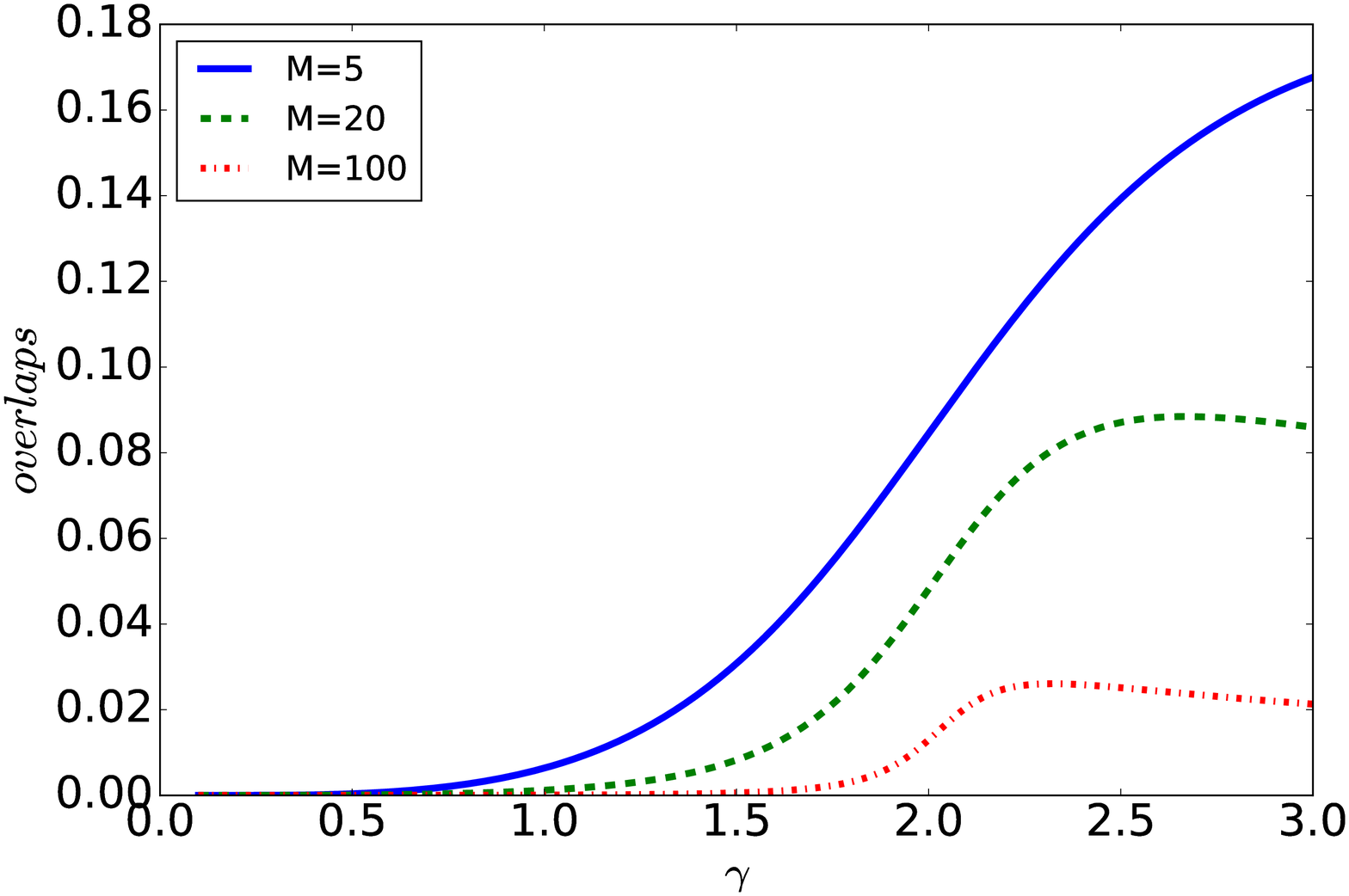}
  \caption{The overlap $|\bra{a}\ket{\psi_1}|^2$ as a function of $\gamma$ for $M=5, 20, 100$.}\label{S03_a_overlap}
\end{figure}

For quantum search on Cayley tree of height 2 with different values of branching factor $M$, the overlap $|\bra{a}\ket{\psi_1}|^2$  is shown as a function of the jumping rate $\gamma$ in Fig.\ref{S03_a_overlap}. The value $|\bra{a}\ket{\psi_1}|^2$ gets smaller when $M$ increases. This explains the decreasing of success probability for the scheme proposed in \cite{agliari2010quantum} when the branching factor $M$ increases.

\section{Probability of success for our single-stage scheme on a weighted Cayley tree of Height 2}\label{success weighted}

\begin{figure}[bt]
\includegraphics[width=0.44\textwidth]{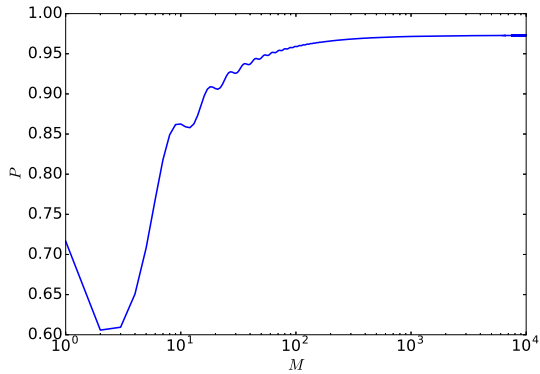}
\caption{Probability of success $P$ versus the branching factor $M$ for our single-stage scheme on a Cayley tree of height $2$ with adjusted weight $\omega=M$ and jumping rate $\gamma=1+1/M$.}
\label{S04_M_success_weighted_tree_height2}
\end{figure}

For our single-stage quantum search scheme on a Cayley tree of height 2, with the adjusted weight $\omega=M$ and the jumping rate $\gamma=1+1/M$, the probability of success as a function of the branching factor $M$ is shown in Fig.\ref{S04_M_success_weighted_tree_height2}. We see that the success probability is close to $100\%$ when $M$ is large. The deviation from $100\%$ might be caused by the deviation of $\gamma$ we chose from the critical jumping rate.

\section{Our single-stage scheme on a Cayley tree of height 3}\label{merging height 3}

\begin{figure}[bt]
\includegraphics[width=0.44\textwidth]{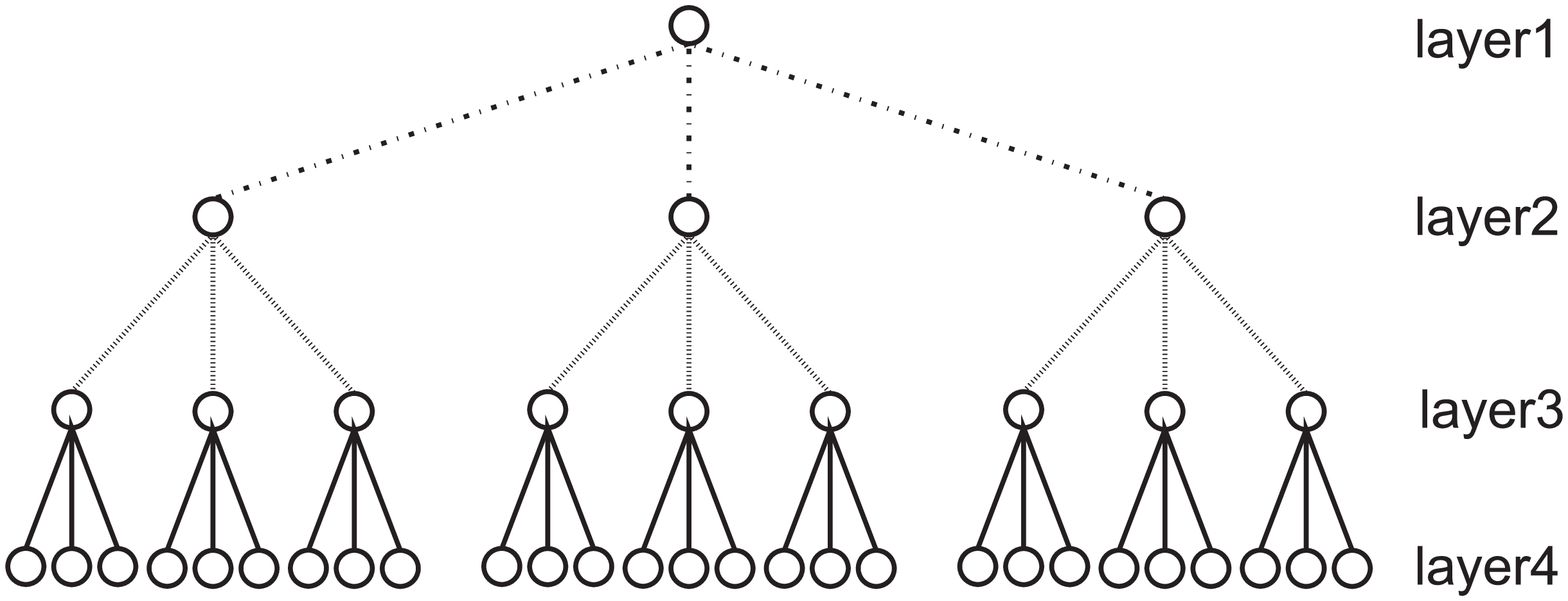}
\caption{An example of weighted Cayley tree of height $3$. The weight of edges between the first and second layer is $\omega_2$, and the weight of edges between the second and third layer is $\omega_1$.}
\label{S05_2nd_tree_weighted}
\end{figure}

The merging of stages could also happen on the weighted Cayley tree of height $3$. As shown in Figure \ref{S05_2nd_tree_weighted}, the weight between the first and second layer is set as $\omega_2$, the weight between the second and third layer is set as $\omega_1$. Numerical calculations show that when $\omega_2$ is larger than roughly $O(M^{\frac{3}{4}})$ and $\omega_1=1$, the number of stages becomes two; when $\omega_1$ is larger than roughly $O(M^{\frac{3}{4}})$ and $\omega_2=1$, the number of stages also becomes two; when $\omega_1$ is larger than roughly $O(M^{\frac{3}{4}})$ and $\omega_2$ is larger than roughly $O(M^{\frac{3}{2}})$, then the number of stages becomes one. In order to have a single stage process, we choose $\omega_1=M$ and $\omega_2=M^2$.
Numerical calculations show that when $\gamma=1+\frac{1}{M}$, we have $E_{1}-E_{0}=2M^{-\frac{3}{2}}$, and time consumed in the search process is $t=\pi M^{\frac{3}{2}}/2 \propto N^{\frac{1}{2}}$. Success probability is larger than $99\%$. Compared to the original three-stage search process which consumes time $t\propto M^{\frac{5}{2}} \propto N^{\frac{5}{6}}$, a substantial speedup is achieved.

\section{Weighted Cayley trees with a small branching factor $M$}\label{overlap small M}

\begin{figure}[bt]
\includegraphics[width=0.44\textwidth]{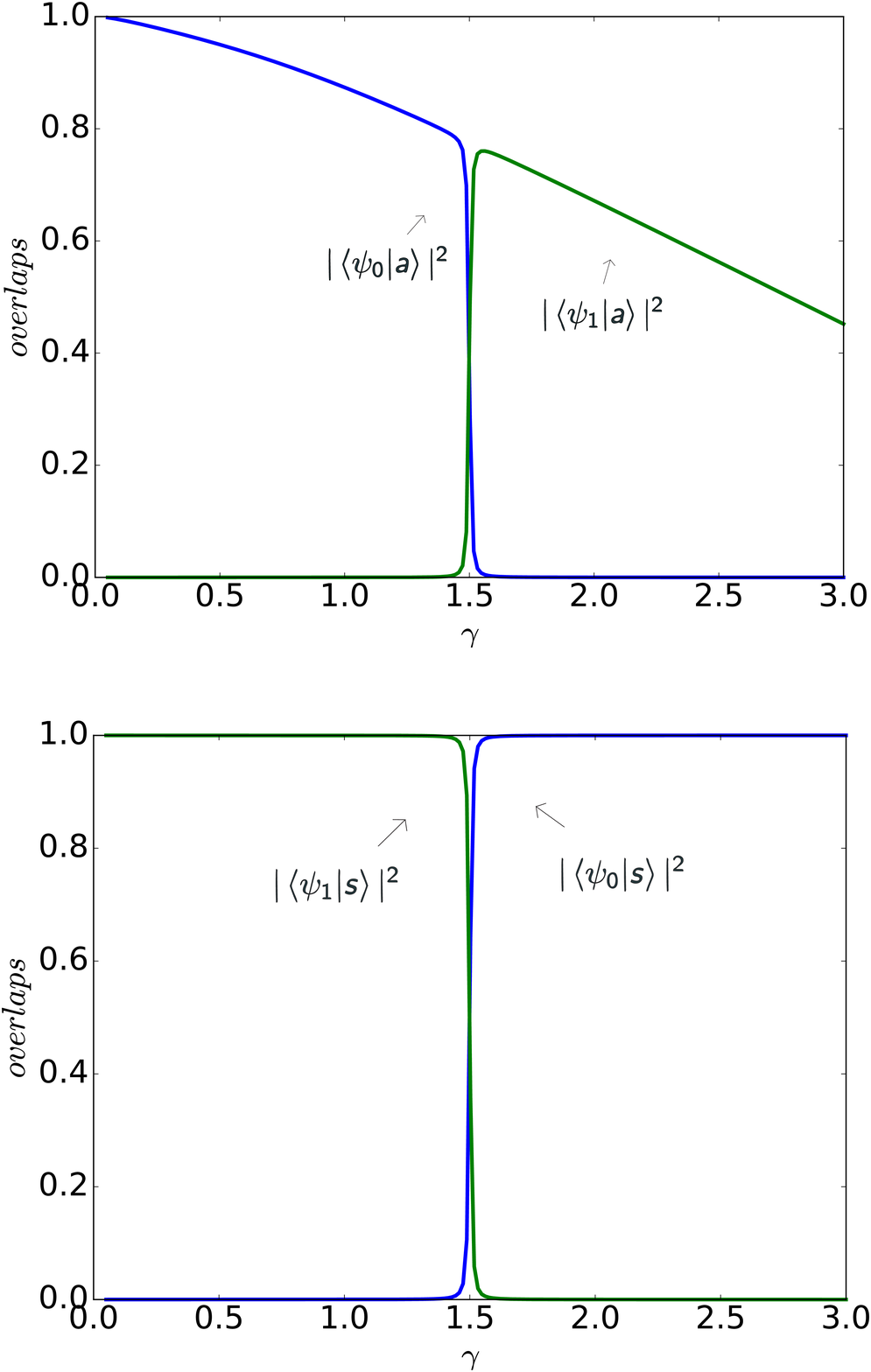}
\caption{The squared overlaps of basis states with the eigenstates of $H$ for weighted Cayley tree of height $15$, $M=2$, and $\omega=3$.}
\label{S06_overlap_balanced_tree_omega_3}
\end{figure}

The squared overlaps of basis states with the eigenstates of $H$ for a weighted Cayley tree with height $15$, a small branching factor $M=2$, and $\omega=3$ are shown in Fig. \ref{S06_overlap_balanced_tree_omega_3}. For $\gamma=1.5$, the ground state $\ket{\psi_0}$ and the first excited state $\ket{\psi_1}$ dominate $\ket{a}$ and $\ket{s}$, hence the search process requires only one stage.

\section{Success Probability of our single-stage scheme on a Weighted Cayley tree with small $M$}\label{success small M}

\begin{figure}[bt]
\includegraphics[width=0.44\textwidth]{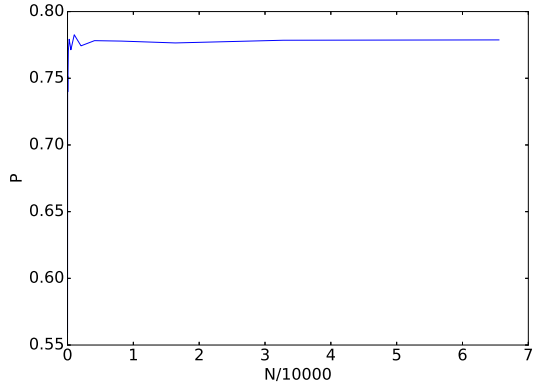}
\caption{The effect of total number of vertices $N$ on the success probability $P$ for weighted Cayley tree of $M=2$, $\omega=3$.}
\label{S07_balanced_tree_success}
\end{figure}

For our single-stage quantum search scheme on a weighted Cayley tree with a small branching factor $M=2$ and a weight $\omega=3$, when $\gamma=1.5$, the probability of success as a function of the total number of vertices is shown in Fig.\ref{S07_balanced_tree_success}. The deviation of success probability from unity is determined by the spectrum of the Hamiltonian $H$ as shown in Sec. \ref{overlap small M}. More suitable weight of Cayley tree might support quantum search with success probability even closer to $100\%$.

\section{connectivity}\label{connectivity}

A Cayley tree has a low connectivity, irrespective of which measure one uses. There are several measures of connectivity \cite{PhysRevLett.114.110503,beineke2002average,chung1996spectral,fiedler1973algebraic}. In Table \ref{connectivity table}, we summarize the connectivity measures with different definitions for various Cayley trees, the first-order truncated $M$-simplex lattice and joined complete graphs for comparison. The average connectivity of the first-order truncated $M$-simplex lattice is  roughly $1/\sqrt{N}$ and the average connectivity of joined complete graphs is roughly $1/2$. Joined complete graphs were used as examples of poorly connected graphs to show connectivity is not a reliable indicator of faster quantum search in \cite{PhysRevLett.114.110503}. Here we provide a better example. Our scheme shows that optimal quantum search can still be achieved on Cayley trees that has an even lower connectivity.  Cayley trees are intrinsically and much more poorly connected as they become disconnected whichever edge is removed.

\linespread{1.5}
\begin{table*}
\centering \caption{Connectivity measures with different definitions, where $r$ stands for the height of Cayley tree, $N$ the total number of vertices, $\omega$ the weight of edge discussed in the main text. These results are obtained by approximation. \label{connectivity table}}
\begin{tabular}{|c|c|c|c|c|}
\hline
Graph & Vertex or edge & Algebraic & Normalized algebraic & Average\\
\hline
Cayley tree $r=2$, $\omega=1$ & 1 & $1/\sqrt{N}$ & $1/2\sqrt{N}$ & $2/N$\\
\hline
Cayley tree $r=2$, $\omega=\sqrt{N}$ & 1 & 0.5 & 0.293 & $4/N$\\
\hline
Cayley tree $r=3$, $\omega_1=1$, $\omega_2=1$ & 1 & $1/N^\frac{2}{3}$ & $1/2N^\frac{2}{3}$ & $2/N$\\
\hline
Cayley tree $r=3$, $\omega_1=N^{\frac{1}{3}}$, $\omega_2=N^{\frac{2}{3}}$ & 1 & 0.332 & 0.134 & $6/N$\\
\hline
Joined Complete & 1 & 1 & $1/N$ & $1/2$\\
\hline
Simplex Complete & $\sqrt{N}$ & 1 & $1/\sqrt{N}$ & $1/\sqrt{N}$\\
\hline
\end{tabular}
\end{table*}

\section{Robustness of quantum search on Cayley trees}\label{robustness}

\subsection{Deviation of jumping rate $\gamma$ from $\gamma _{c}$}

We have mentioned when $\gamma$ is considerably away from $\gamma _{c}$, the system state would essentially evolve only by a phase factor.
Here we discuss the case when $\gamma$ deviates from $\gamma _{c}$ by a small amount, and try to find how the deviation affects the quantum search.
We find that a relatively small deviation of the jumping rate decreases the probability of success as we expect, however, it makes the quantum search faster!

As discussed in \cite{J.Phys.A:Math.Theor.49.484002}, with potential barriers, the amplitude hopping from any vertex to other vertices is decreased by a factor $\epsilon$, the Laplacian of the graph $L$ becomes $L'=L(1-\epsilon)$. This is equivalent to the deviation of jumping rate $\gamma$ since $H=-\gamma L'-\ket{\omega}\bra{\omega}=-\gamma'L-\ket{\omega}\bra{\omega}$ with $\gamma'=(1-\epsilon)\gamma$.

For Cayley tree of height $3$, when we choose $\omega_1=M$, $\omega_2=M^2$ and $\gamma=1$, we find that time consumed in the search process is $t=\pi M$. Since the number of vertices is $N=M^3+M^2+M+1$, we have $t\propto N^{\frac{1}{3}}$, which is even faster than Grover's algorithm. Since the critical jumping rate is $\gamma _{c}=1+\frac{1}{M}$, it seems faster quantum search is achieved when $\gamma$ deviates from $\gamma _{c}$. However, we also find the success probability is $20\%$ for $M=100$, $9\%$ for $M=500$, and $6.5\%$ for $M=1000$. (In previous discussion, success probability is independent of $M$, as long as $M$ is larger enough.)

We would like to point out that when $\gamma=\gamma _{c}$, the energy gap $\Delta E$ between the two related eigenstates is the smallest. Since $t\propto  \Delta E^{-1}$, when energy gap becomes larger, time consumed in the search process will decrease together with a decrease of the success probability. But we find that the expected average time consumed for the search to succeed differs by only a constant factor for a small deviation of $\gamma$. Take Cayley tree of height $3$ as an example, when $\omega_1=M$, $\omega_2=M^2$, $\gamma=1+\frac{1}{M}\pm \frac{1}{M}$, numerically we find that time consumed is $t\propto M$, success probability is $p\propto M^{-\frac{1}{2}}$,  and $p/t\propto M^{-\frac{3}{2}}$. When $\omega_1=M$, $\omega_2=M^2$, $\gamma=1+\frac{1}{M}$, success probability is $p_0\approx 1$, time consumed is $t_0\propto M^{\frac{3}{2}}$, $p_0/t_0\propto M^{-\frac{3}{2}}$. The expected average time consumed is $<t_{success}>=pt+2p(1-p)t+3p(1-p)^2t+\cdots=t/p$. Thus the expected average time consumed for the two search process differ by only a constant factor for Cayley trees of height $3$. We have checked that similar results are also obtained for Cayley trees of other height. It is not clear how large the deviation can be without ruining the quantum search. When $\gamma$ is far from $\gamma_c$, whether the expected average time consumed still differs by only a constant factor is an open question.

\subsection{Small perturbation on graph structure}

We perform perturbation on the structure of the graph, and discuss how the modification affects the two-stage quantum search when $M$ is large enough. We obtain the following results.

\begin{figure}[bt]
\includegraphics[width=0.44\textwidth]{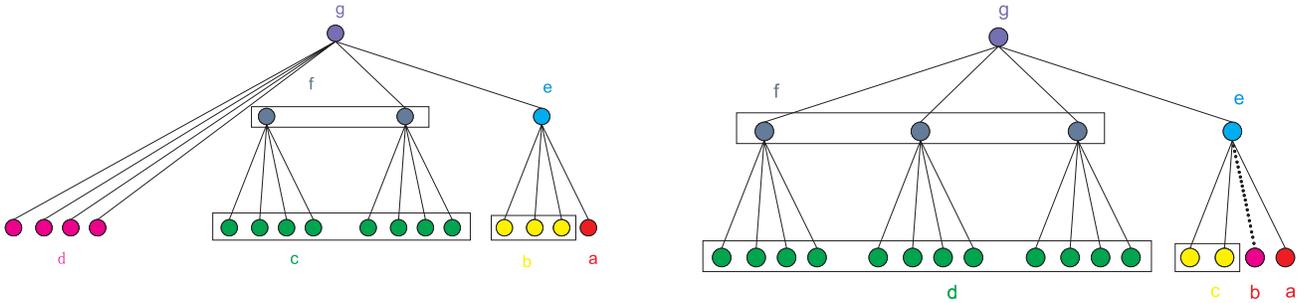}
\caption{Perturbation I: a group of bottom vertices labeled with $d$ are directly connected to the top of the tree.}
\label{S08_tree_small_1}
\end{figure}

(1) As shown in Figure \ref{S08_tree_small_1}, the perturbation connects group of bottom vertices labeled with $d$ directly to the top of the tree. No large influence on the two-stage search is observed. The quantum search is quite robust with respect to this kind of perturbation.

\begin{figure}[bt]
\includegraphics[width=0.44\textwidth]{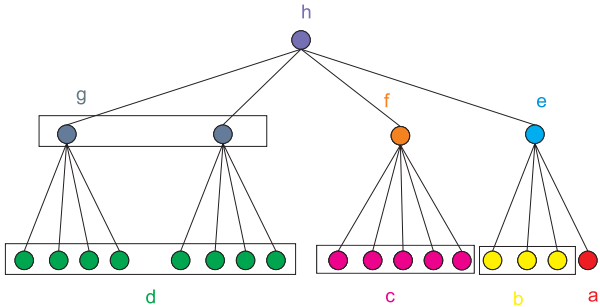}
\caption{Perturbation II: the number of vertices in one subgroup $c$ is changed form $M$ to $m$}
\label{S09_tree_small_2}
\end{figure}

(2) As shown in Figure \ref{S09_tree_small_2}, we change the number of vertices in one group $c$ from $M$ to $m$. When $m$ is not too large, i.e., $m=o(M^{\frac{3}{2}})$, the two-stage search succeeds with success probability larger than $99\%$. Quite robust!

\begin{figure}[bt]

  \includegraphics[width=0.44\textwidth]{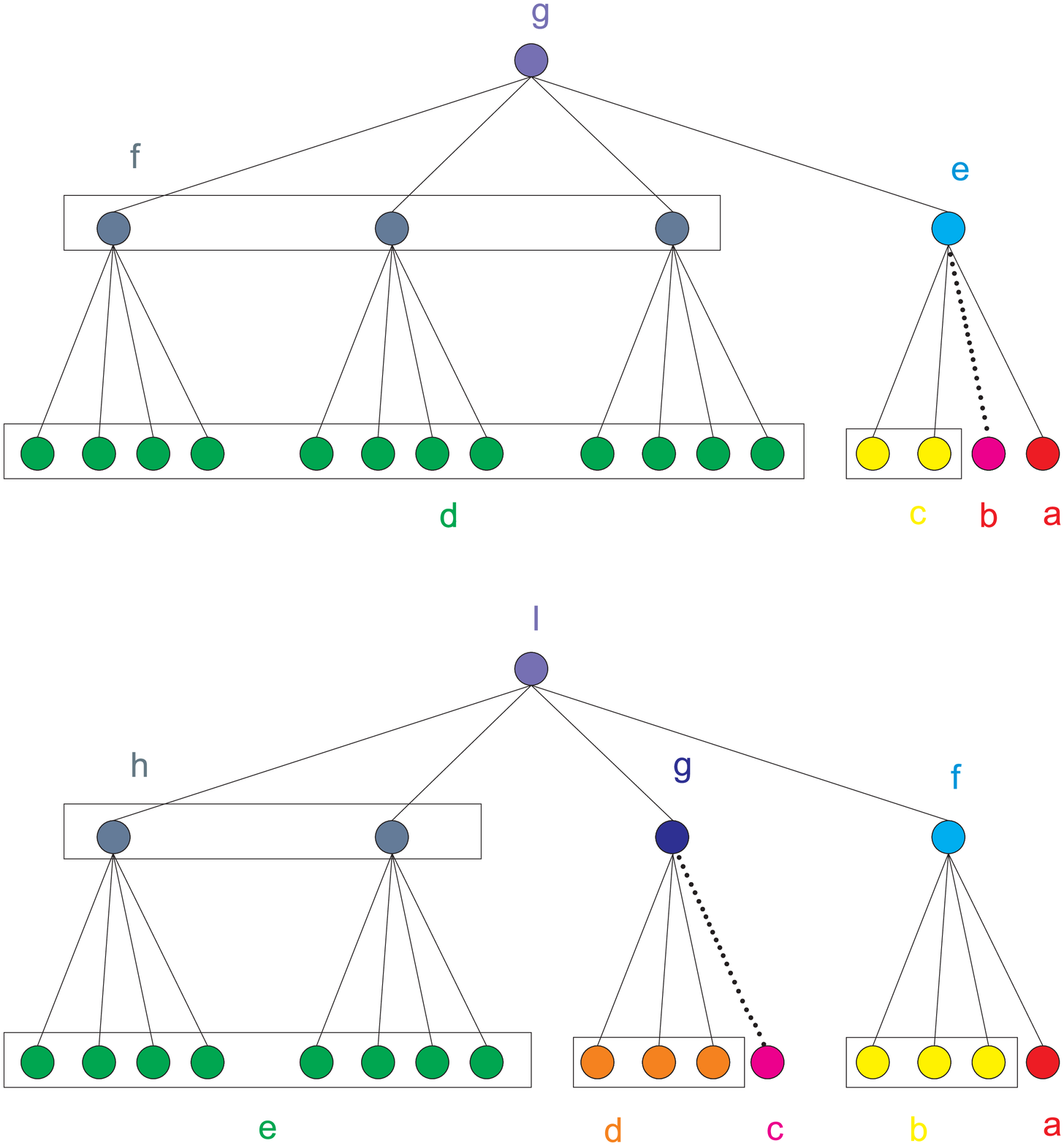}

  \caption{Perturbation III: the weight of one edge is changed. Dotted edges have weight $\omega$}\label{S10_tree_small_3}
\end{figure}

(3) As shown in Figure \ref{S10_tree_small_3}, we change the weight of one edge between two vertices in groups $\ket{b}$ and $\ket{e}$ or the weight of one edge between two vertices in groups $c$ and $g$ to $\omega$. When $\omega$ is not too large, i.e., $\omega=o(M^3)$, no influence on the two-stage search is observed. Quite robust!

\begin{figure}[bt]
  \includegraphics[width=0.45\textwidth]{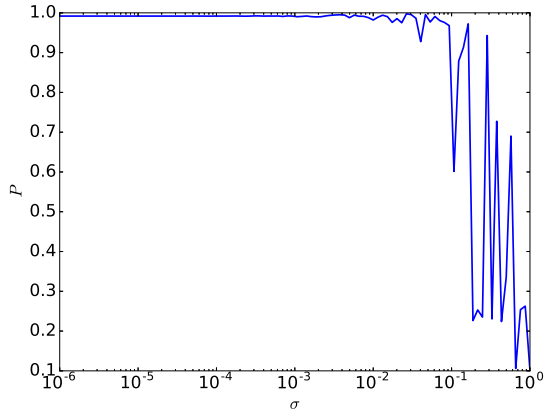}
  \caption{The influence of Gaussian noise on the success probability for a Cayley tree of height 2 with $M=33$}\label{S11_sigma_success_tree_height2}
\end{figure}

(4) Introducing Gaussian noise to every edge of a Cayley tree of height 2, the influence of noise on the success probability is shown in Figure \ref{S11_sigma_success_tree_height2}. When standard deviation $\sigma$ is smaller than $10^{-2}$, the success probability is not strongly affected.

\begin{figure}[bt]
  \includegraphics[width=0.44\textwidth]{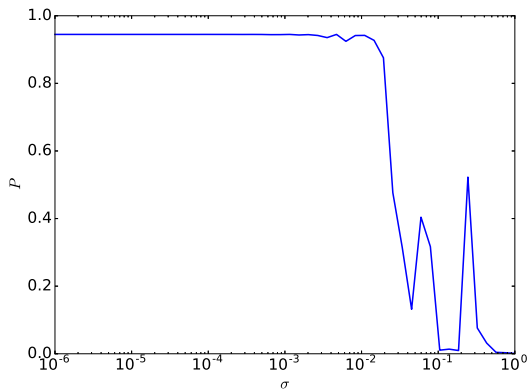}
  \caption{The influence of Gaussian noise on the success probability for a weighted Cayley tree of height 2 with $M=50$}\label{S12_sigma_success_weighted_tree_height2}
\end{figure}

(5) Introducing Gaussian noise to every edge of a weighted Cayley tree of height 2, the influence of noise on the success probability is shown in Figure \ref{S12_sigma_success_weighted_tree_height2}. When standard deviation $\sigma$ is smaller than $10^{-2}$, the influence of noise is very weak.

Therefore, our quantum search schemes, the multi-stage scheme as well as the single-stage scheme with adjustable weights, are quite robust under small perturbation of the graph structure.

\subsection{Large modification on graph structure}

Now we consider large modifications on graph structure.

\begin{figure}[bt]
\includegraphics[width=0.44\textwidth]{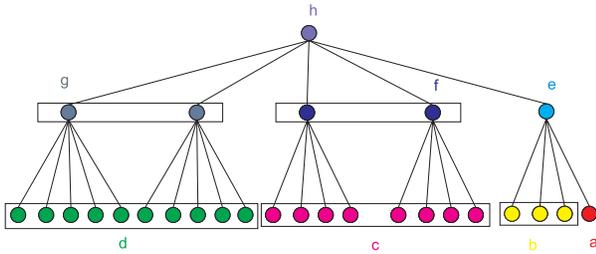}
\caption{Large modification I: the number of vertices in nearly half of the groups is changed.}
\label{S13_tree_large_1}
\end{figure}

(1) As shown in Figure \ref{S13_tree_large_1}, we change the number of vertices in nearly half of groups in the ground layer to $m$. When $m=o(M^{\frac{3}{2}})$, the success probability of the two-stage search process is larger than $99\%$. However, time consumed in each stages would change by a constant.

\begin{figure}[bt]
\includegraphics[width=0.44\textwidth]{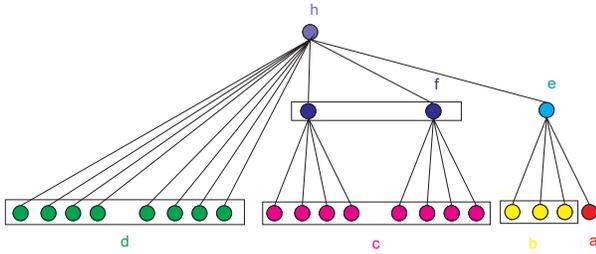}
\caption{Large modification II: nearly half of the groups in the bottom layers are directly connected to the top of the tree.}
\label{S14_tree_large_2}
\end{figure}

(2) As shown in Figure \ref{S14_tree_large_2}, nearly half of the groups in the bottom layer are directly connected to the top of the tree. The search process fails in such a modified graph.

(3) For Cayley tree of height 2, when each edge is assigned the weight $\omega=1$ or $\omega=2$ with equal probability, the search process is not successful.

A large modification on graph structure could affect our quantum search considerably.

\end{document}